\pdfoutput=1

\documentclass[11pt]{article}

\usepackage[final]{acl}

\usepackage{times}
\usepackage{latexsym}

\usepackage[T1]{fontenc}

\usepackage[utf8]{inputenc}

\usepackage{microtype}

\usepackage{inconsolata}

\usepackage{graphicx}

\title{From Knowledge to Noise: CTIM-Rover and the Pitfalls
of Episodic Memory in Software Engineering Agents}

\author{Tobias Lindenbauer\\
  School of Computation, Information, and Technology, Technical University of Munich, Germany \\
  \texttt{tobias.lindenbauer@tum.de}}

\author{
\textbf{Tobias Lindenbauer\textsuperscript{1}},
\textbf{Georg Groh\textsuperscript{1}},
\textbf{Hinrich Schütze\textsuperscript{2}}
\\
\textsuperscript{1}School of Computation, Information and Technology, Technical University of Munich\\
\textsuperscript{2}CIS \& MCML, LMU Munich \\
\small{
\textbf{Correspondence:} \href{mailto:tobias.lindenbauer@tum.com}{tobias.lindenbauer@tum.com}
  }
}

\usepackage[nolist]{acronym}
\begin{acronym}
\acro{llm}[LLM]{Large Language Model}
\acro{icl}[ICL]{In-Context Learning}
\acro{swe}[SE]{Software Engineering}
\acro{cl}[CL]{Continual Learning}
\acro{nlp}[NLP]{Natural Language Processing}
\acro{ctim}[CTIM]{Cross-Task-Instance Memory}
\acro{moe}[MoE]{Mixture-Of-Expert}
\acro{kd}[KD]{Knowledge Distillation}
\acro{kb}[KB]{Knowledge Base}
\acro{cot}[CoT]{Chain-of-Thought}
\acro{el}[EL]{Experiential Learning}
\end{acronym}
\usepackage{amssymb}
\usepackage{mathtools}
\usepackage{tcolorbox}
\usepackage{multicol,multirow}

\begin{document}
\maketitle
\begin{abstract}
We introduce CTIM-Rover\footnote{\url{https://github.com/Liqs-v2/ctim-rover}}, an AI agent for \ac{swe} built on top of AutoCodeRover~\cite{zhang_autocoderover_2024} that extends agentic reasoning frameworks with an episodic memory, more specifically, a general and repository-level \emph{\ac{ctim}}. While existing open-source \ac{swe} agents mostly rely on ReAct~\cite{yao_react_2023}, Reflexion~\cite{shinn_reflexion_2023}, or CodeAct~\cite{wang_executable_2024}, all of these reasoning and planning frameworks inefficiently discard their long-term memory after a single task instance. As repository-level understanding is pivotal for identifying all locations requiring a patch for fixing a bug, we hypothesize that \ac{swe} is particularly well positioned to benefit from \ac{ctim}. For this, we build on the \ac{el} approach ExpeL~\cite{zhao_expel_2024}, proposing a \acfp{moe} inspired approach to create both a general-purpose and repository-level \ac{ctim}. We find that \ac{ctim}-Rover does not outperform AutoCodeRover in any configuration and thus conclude that neither ExpeL nor DoT-Bank~\cite{lingam_enhancing_2024} scale to real-world \ac{swe} problems. Our analysis indicates noise introduced by distracting \ac{ctim} items or exemplar trajectories as the likely source of the performance degradation.
\end{abstract}

\section{Introduction}
\label{sec:introduction}
AI Agents have recently proven themselves as a competitive way of scaling test-time compute, especially in \ac{swe}~\cite{openai_swe_bench_verified_2024}. A crucial yet underexplored component of AI agents is their memory, which allows them to dynamically adapt their behavior based on prior experiences. Early approaches, such as ReAct~\cite{yao_react_2023}, rely on the agent’s immediate trajectory or short-term memory for decision-making. Reflexion~\cite{shinn_reflexion_2023} extends this by introducing long-term memory in the form of self-reflections on past failed task attempts, enabling agents to improve their reasoning and planning on a single task instance through~\ac{icl}. While this yields performance gains on the current task instance, Reflexion discards these self-reflections after task completion. This results in inefficient use of computational resources and loss of valuable cross-task-instance learning opportunities. \citet{zhao_expel_2024} address this limitation through \acf{el}, which is learning from past experiences across task instances. Their approach ExpeL achieves promising results on HotpotQA~\cite{yang_hotpotqa_2018}, WebShop~\cite{yao_webshop_2023}, and Alfworld~\cite{shridhar_alfworld_2021}. To better align with existing terminology, we name the memory consisting of knowledge extracted with \ac{el} ``\ac{ctim}''. Our work investigates whether \ac{ctim} generalizes to the more complex\footnote{\acs{ctim}-Rover's mean context is $\approx4$ times larger than ExpeL's on HotpotQA~\cite{yang_hotpotqa_2018}. Details in Table~\ref{tab:appendix:trajectory-stats}.} domain of \ac{swe}. We choose \ac{swe} because we expect \ac{el} to be particularly valuable for uncovering the structure of a repository, reducing the number of turns taken exploring the codebase.

\begin{figure*}[ht]
    \centering
    \includegraphics[width=\textwidth]{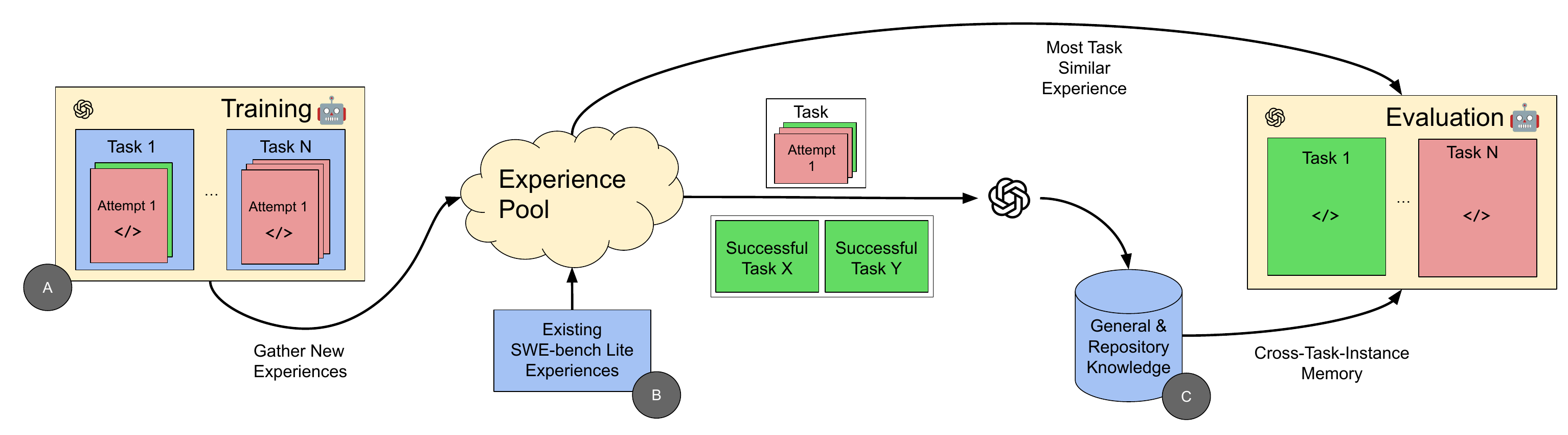}
    \caption{\textbf{\acs{ctim}-Rover Overview.} Figure inspired by ExpeL~\cite{zhao_expel_2024}. \acs{ctim}-Rover first gathers new experiences on the train set of SWE-bench Verified which we introduce in Section~\ref{sec:dataset} (details in Appendix~\ref{sec:appendix:dataset}). Then, it combines these experiences with existing experiences of AutoCodeRover~\cite{zhang_autocoderover_2024} on SWE-bench Lite~\cite{jimenez_swe-bench_2023}. Next, it distills high-level and repository-level knowledge from these experiences. During evaluation, it recalls a past experience and conditions on the distilled knowledge. \textbf{Key departures from ExpeL or AutoCodeRover in blue:} (A) We extend AutoCodeRover with Reflexion~\cite{shinn_reflexion_2023}, allowing the agent to retry an instance up to three times while learning from its mistakes through self-reflection.  (B) Compared to ExpeL, we also source experiences from past successful trajectories outside our system. (C) We introduce a novel domain-specific \ac{kd} phase (Figure~\ref{fig:repo-distill}) that extracts repository-level insights (e.g., common bug patterns).}
    \label{fig:core-contributor-system}
\end{figure*}

To adapt \ac{el} to \ac{swe} we extend it to a \acp{moe} inspired \acf{kd} approach that simultaneously captures high-level \ac{swe} best practices and repository-specific details (e.g., project structure). We experimentally evaluate this approach by augmenting AutoCodeRover~\cite{zhang_autocoderover_2024} with \ac{ctim}, which we name ``\acs{ctim}-Rover'', and comparing the results of \acs{ctim}-Rover with those of the AutoCodeRover on a subset of SWE-bench Verified. We find that our adapted \ac{ctim} does not generalize to \ac{swe} and instead degrades performance in all configurations compared to AutoCodeRover. Our detailed qualitative analysis identifies noisy \ac{ctim} items as culprits and we propose the use of embedding-based retrieval methods to provide relevant, task-similar \acp{ctim} items. The potential of this approach in the \ac{swe} domain was recently demonstrated by~\cite{su_learn-by-interact_2025} who provided relevant sub-trajectories for \ac{icl} at each agent turn.

\section{Related Work}
\label{sec:related-work}
\subsection{Agentic Reasoning Frameworks}
\label{subsec:agentic-reasoning-frameworks}
A core element of popular agentic reasoning frameworks~\cite{yao_react_2023,shinn_reflexion_2023,wang_executable_2024} is the agent's trajectory or short-term memory, consisting of its past actions, reasoning and environment observations. \citet{shinn_reflexion_2023} introduce a long-term memory consisting of self-reflections over the short-term memory of unsuccessful previous attempts. However, after concluding a task instance, existing reasoning frameworks used in \ac{swe} agents do not further use the short- or long-term memory. Our work addresses this key limitation by adapting ExpeL~\cite{zhao_expel_2024} to the \ac{swe} domain.

\subsection{\ac{swe} Agents}
\label{subsec:swe-agents}
SWE-agent~\cite{yang_swe-agent_2024} was the first openly available \ac{swe} agent and leverages the ReAct reasoning framework~\cite{yao_react_2023}. The agent's basic search tooling combined with its interleaved bug localization and patch generation approach offers flexibility, but results in long and expensive trajectories. AutoCodeRover~\cite{zhang_autocoderover_2024} on the other hand, explicitly structures the task into two distinct phases: bug localization and patch generation. Additionally, it provides sophisticated search tooling during localization and constrains the patch generation phase to a maximum of three retry attempts. This ensures shorter, cost-efficient trajectories and a guaranteed termination shortly after the patch generation step. A key limitation of this approach is that the agent cannot gather additional context once it enters the patch generation phase. However, current \ac{swe} agents are not yet capable of recovering from early mistakes, and their performance stagnates at later turns~\cite{yang_swe-smith_2025}. Furthermore, neither of these agents employ \ac{ctim}. Thus, our work expands the cost-efficient AutoCodeRover with \ac{ctim}.

\subsection{Concurrent Work}
\citet{lingam_enhancing_2024} perform self-reflection on the same task instance while prompting for a diverse set of self-reflections and additionally enhance the context with exemplar trajectories from other task instances. This approach demonstrates performance gains on programming benchmarks with comparatively short trajectories (e.g., HumanEval~\cite{chen_evaluating_2021}). Especially the latter setup is closely related to \ac{ctim}-Rover with an exemplar trajectory. However, we evaluate on SWE-bench~\cite{jimenez_swe-bench_2023} which more closely resembles real \ac{swe} tasks. Instead of abstracting from the trajectory by constructing a \ac{ctim}, ~\cite{su_learn-by-interact_2025} directly retrieve synthetic sub-trajectories at each step of the agent and achieve strong performance on SWE-bench. Furthermore, we provide the full \ac{ctim} with the user prompt at the start of an agent's trajectory instead of select subset at each turn.

\begin{figure*}[th]
    \centering
    \includegraphics[width=0.9\textwidth]{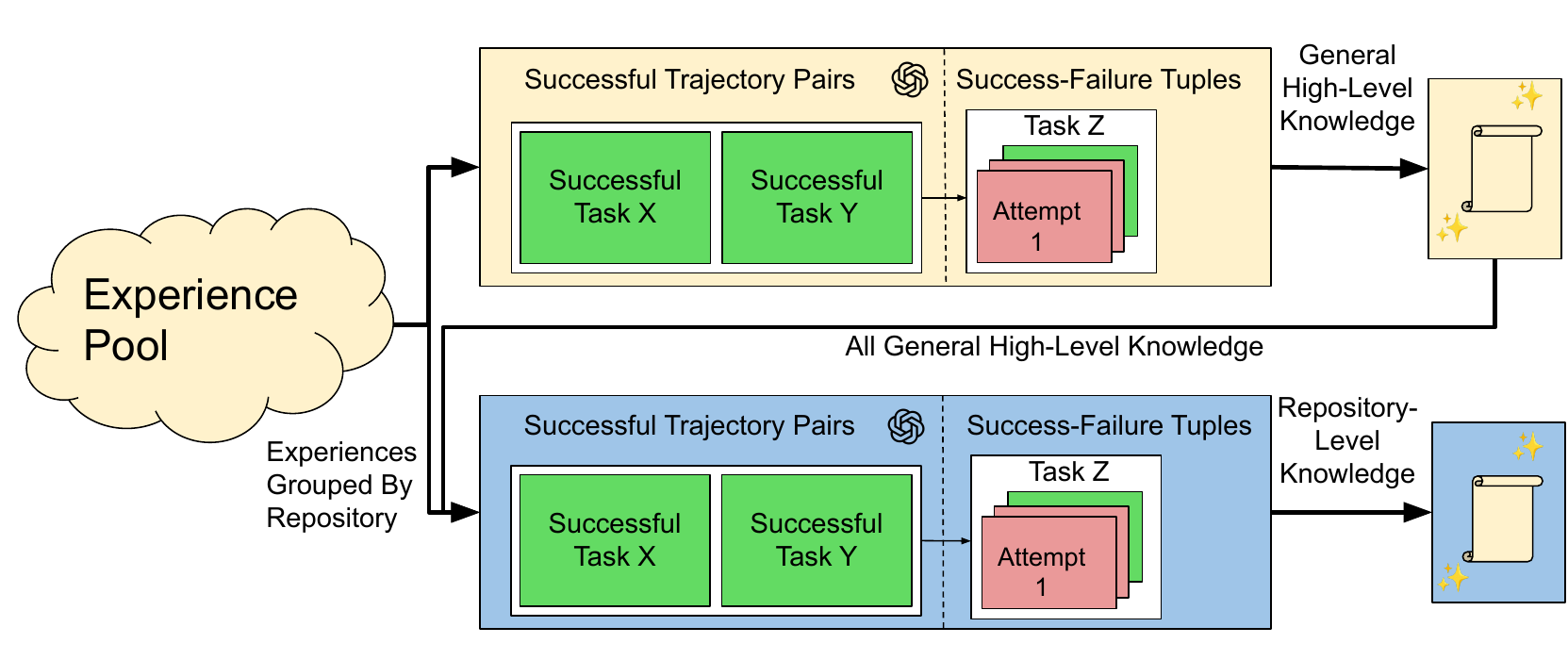}
    \caption{\textbf{CITM-Rover \acf{kd}. Key departure from ExpeL~\cite{zhao_expel_2024} in blue.} \textbf{Top:} (1) Distill generally applicable \ac{swe} knowledge from pairs of successful trajectories from different task instances and (2) tuples of a successful task instance and its self-reflection retries. \textbf{Bottom:} (3) 
    Use the generally applicable knowledge and past experience to distill repository-level knowledge from pairs of successful trajectories from different task instances within the same repository and (4) tuples of a successful task instance and its self-reflection retries for a given repository.}
    \label{fig:repo-distill}
\end{figure*}

\section{Dataset}
\label{sec:dataset}
We use SWE-bench Verified~\cite{openai_swe_bench_verified_2024} without samples from the \texttt{pylint}, \texttt{astropy} and \texttt{pydata/xarray} repositories due to environment setup issues~\footnote{Thanks to the AutoCodeRover authors for helping us validate these.} as basis for our experiments. For details see Section~\ref{sec:limitations}. For our experiments, we rely on SWE-bench Verified, opposed to SWE-bench~\cite{jimenez_swe-bench_2023}, as it guarantees that samples are theoretically solvable~\cite{openai_swe_bench_verified_2024}. For the collection of past successful trajectories (Section~\ref{subsec:swe-bench:train-collection}) we use 401 samples from this benchmark and for the evaluation 45 samples.

\subsection{Systematic Collection of Past Successful Trajectories}
\label{subsec:swe-bench:train-collection}
To construct a high quality \ac{ctim}, we require a diverse and representative set of successful past trajectories. These are past experiences on SWE-bench in which the agent solved an instance. This section details our systematic approach to collecting these trajectories.

To generate as many successful past trajectories as possible, we extend the baseline AutoCodeRover~\cite{zhang_autocoderover_2024} implementation with self-reflection capabilities. Following~\citet{shinn_reflexion_2023}, we retry an instance up to three times and allow self-reflections to inform each subsequent attempt. While AutoCodeRover allows up to three patch generation attempts, this does not entail a complete retry on the full trajectory, nor a self-reflection between the patch generation attempts. During training we reduce the patch generation retries of AutoCodeRover from three to two to amortize some of the additional cost incurred by Reflexion retries. With this setup we gather the trajectories of 183 successfully solved instances. To further increase our training set, we supplement the collected trajectories with 53 successful AutoCodeRover trajectories from SWE-bench Lite. Because \acs{ctim}-Rover's trajectories only differ from vanilla AutoCodeRover trajectories by the addition of self-reflections, and both SWE-bench Verified and SWE-bench Lite are subsets of SWE-bench we consider this operation valid with respect to our data distribution. We use these 236 past successful trajectories to construct our \ac{ctim}. For details on their distribution see Appendix~\ref{sec:appendix:training-trajectory-distr}.

\begin{table*}[t]
\centering
\small
\begin{tabular}{l|ccccccc|c}
\hline
\textbf{Configuration} & \textbf{Django} & \textbf{Matplotlib} & \textbf{Mwaskom} & \textbf{Pytest} & \textbf{Scikit} & \textbf{Sphinx} & \textbf{Sympy} & \textbf{Overall} \\
 & \textbf{(22)} & \textbf{(4)} & \textbf{(1)} & \textbf{(3)} & \textbf{(1)} & \textbf{(5)} & \textbf{(9)} & \textbf{(45)} \\
\hline
AutoCodeRover & 50 & 25 & 0 & 33 & \textbf{100} & 0 & \textbf{56} & \textbf{42} \\
\acs{ctim}-Rover & 50 & \textbf{50} & 0 & 33 & \textbf{100} & 0 & 33 & 40 \\
~~~\acs{ctim} only & 36 & 25 & 0 & 33 & \textbf{100} & 0 & 33 & 31 \\
~~~General \acs{ctim} only & \textbf{55} & 0 & 0 & 33 & \textbf{100} & 0 & 22 & 36 \\
~~~Repo-level \acs{ctim} only & 41 & 0 & 0 & 33 & \textbf{100} & 0 & 33 & 31 \\
~~~Exemplar only & 50 & 25 & 0 & \textbf{67} & \textbf{100} & 0 & 33 & 40 \\
\hline
\end{tabular}
\caption{Success rates (\%) on our test set across \acs{ctim}-Rover configurations and repositories. Values in parentheses indicate the number of samples in our test set per repository.}
\label{tab:success-rates}
\end{table*}

\section{Experiments}
\label{sec:experiments}
To adapt \ac{el} to \ac{swe}, we extend the \ac{ctim} with a \ac{moe}~\cite{jacobs_mixture_1991} inspired repository-level \ac{ctim} (Section~\ref{subsec:experiments:cross-task-instance-memory}) and investigate \ac{icl} with successful, task-similar exemplar trajectories (Section~\ref{subsec:experiments:exemplars}). For distilling knowledge from trajectories, we use the reasoning model o1~\cite{openai-o1} because we suspect that its capabilities are beneficial when identifying pivotal agent decisions in complex \ac{swe} agent trajectories (i.e., cause-effect relationships). We use GPT-4o~\cite{openai-gpt-4o} to power the agent during training trajectory collection and the final evaluations due to budget constraints.

\subsection{\acf{ctim}}
\label{subsec:experiments:cross-task-instance-memory}
Our approach shares the core principle of using knowledge extracted from past successful trajectories to guide the agent on future instance with ExpeL~\cite{zhao_expel_2024}. We provide a high-level system overview of in Figure~\ref{fig:core-contributor-system}. To adapt this approach to \ac{swe}, we extract repository-level knowledge following and conditioned on general \ac{swe} knowledge in a two-phase approach detailed below (Figure~\ref{fig:repo-distill}).

\paragraph{Repository-Level Knowledge Distillation}
Our approach re-uses the \ac{kd} methodology (extracting knowledge from sets of successful trajectories from distinct instances and tuples of successful and failing attempts in the same instance) and operations (add, edit, upvote or downvote\footnote{If the importance of a \ac{ctim} item falls below 0, this operation removes that item from the \ac{ctim}.}) introduced by~\citet{zhao_expel_2024} with the following modifications. First, we double the initial importance value of \ac{ctim} items, because we expect longer intervals between instances for which a \ac{ctim} item is applicable. This is motivated by the limited state space of ExpeL's environments, compared to the complexity of real world software repositories. Furthermore, some of our trajectories contain self-reflections. We expect these trajectories to produce especially high-quality \ac{ctim} items when extracting knowledge from tuples of successful and failing attempts in the same instance as they already contain the insights that lead to an eventual resolution. After the first phase of general \ac{ctim} construction, we build a repository-specific \ac{ctim} by constraining all instances shown to the distilling \ac{llm} (see Section~\ref{sec:experiments}) to be from the same repository. Finally, we limit the maximum size of the \ac{ctim} to $c(n) = \lceil\sqrt{n}\rceil$, where $n$ represents the number of available successful trajectories for constructing this \ac{ctim}. With this we aim to iteratively refine the \ac{ctim} to contain a concise set of high-quality insights and avoid degrading the agent's performance with noisy knowledge. For prompts see Appendix~\ref{sec:appendix:prompt-template}, for sample \ac{ctim} items Appendix~\ref{sample-ctim-items}.

Using the repository-level knowledge, we expect the agent will more efficiently explore its environment by re-using knowledge relating to previously explored areas of its environment. This knowledge may provide insights on (1) the structure of the project, (2) entry points or data flow and architectural patterns, (3) coding conventions encountered, (4) common failure modes relating to the application domain of the software (e.g., failure modes for image processing in OpenCV), or (5) common bugs that the agent encountered in that past.

\subsection{Exemplar retrieval}
\label{subsec:experiments:exemplars}
In addition to providing the \ac{ctim} for \ac{icl}, we investigate if \ac{icl} with the most task-similar past successful trajectory improves performance. For this, we construct a Milvus~\cite{2021milvus} index consisting of problem statement embeddings, using Code-T5~\cite{wang_codet5_2021} base as the embedding model. This model's size allows local use and it is trained for language and code, which our problem statements consist of. During evaluation, we retrieve the most task-similar past successful trajectory based on cosine similarity scores with a 90\% threshold. This ensures an exemplar is only shown if a relevant one is available ($\approx62\%$ of samples).

\section{Results}
\label{sec:results}
We evaluate \acs{ctim}-Rover's performance across the configurations listed in Table~\ref{tab:success-rates}. \acs{ctim}-Rover achieves only a 40\% success rate, which is two percent points worse than our baseline AutoCodeRover. Surprisingly, ``Exemplar only'' configuration matches this performance. The ``\ac{ctim} only'' configuration unexpectedly degraded the performance to just 31\%, 11 percent points less than the baseline. Seeing how poorly \acs{ctim}-Rover performed in the ``Repo-level \ac{ctim} only'' configuration, we partially attribute the performance degradation in the ``\ac{ctim} only'' configuration, to the repository-specific \ac{ctim}. Moreover, we observe a performance degradation even for the ``django'' repository, which our train set is heavily skewed towards (Figures~\ref{fig:appendix:repository-distribution} and~\ref{fig:appendix:successful-trajs:repository-distribution}). We expected instances in this repository to disproportionally benefit from the additional repository-level knowledge due to the reasons discussed in Section~\ref{subsec:experiments:cross-task-instance-memory}. Surprisingly the performance is somewhat stable compared to the baseline, even for underrepresented repositories (e.g., Pytest). This suggests source of the observed performance degradation may relate to the \ac{ctim} usage and quality rather than the quantity. We hypothesize that (1) providing all \ac{ctim} items may introduce unexpected noise because we do not filter these items for relevance regarding the instance's context, and (2) our \ac{ctim} optimization constraint leads to an overly smooth, uninformative and thus noisy \ac{ctim}. To diagnose the reasons for the poor performance, we next perform a detailed qualitative investigation of two randomly chosen samples. 

\subsection{Qualitative Performance Degradation Analysis}
We first consider ``django\_\_django-13933'', a sample that our baseline solves, but \acs{ctim}-Rover with ``Repo-level \acs{ctim} only'' does not. Initially, both systems invoke the correct API returning \texttt{to\_python}, the function that needs a patch. However, our system decides to further investigate the \texttt{clean} function, which is also returned by the API, and does not further investigate \texttt{to\_python}. This indicates an unexpected bias towards the tokens constituting ``clean''. In the repository-level \ac{ctim} for ``django'' we notice that the item in Figure~\ref{fig:django-insight} contains the word \textbf{clean}. Upon removing this item from the \ac{ctim} and retrying, our system correctly identifies the \texttt{to\_python} function as the location for the patch and solves the sample. 

\begin{figure}[t]
    \centering
    \begin{tcolorbox}[
        colback=gray!10,
        colframe=gray!50!black,
        arc=0mm,
        boxrule=1pt,
        width=\columnwidth,
        fonttitle=\bfseries,
        title=Problematic ``django'' \ac{ctim} Item
    ]
    \small
    [...] Ensure to separate resolution from the final redirect to keep path\_info \textbf{clean} while preserving the prefix in the final URL, preventing forced [...]
    \end{tcolorbox}
    \caption{Excerpt of the repository-level \ac{ctim} item that biased our system toward investigating the incorrect \texttt{clean} function, demonstrating how seemingly innocuous knowledge can misguide the agent.}
    \label{fig:django-insight}
\end{figure}

Next, we focus on ``django\_\_django-15987'', a sample that both AutoCodeRover and \acs{ctim}-Rover with ``Repo-level \acs{ctim} only'' solved, but \acs{ctim}-Rover failed to solve in the ``\acs{ctim} only'' configuration. The problem statement of this sample explicitly mentions the constant \texttt{FIXTURE\_DIRS} and AutoCodeRover correctly searches the repository for this constant. However, \acs{ctim}-Rover with the ``\acs{ctim} only'' configuration does not. We notice that our \ac{ctim} does not refer to any constants and suspect that this biases our system towards lower snake-case names. Upon adding the arbitrary, capitalized item ``GRANDMA LIKES PASTA'' to the \ac{ctim} and retrying, our system again solves the sample. This suggests noisy \ac{ctim} biases \acs{ctim}-Rover toward suboptimal initial steps rather than helping it skip initial exploration turns and furthermore hypothesize that lengthy exemplar trajectories likely cause similar issues.

\section{Conclusion}
\label{sec:conclusion}
In this work we extend ExpeL~\cite{zhao_expel_2024}, which showed
promising performance on HotpotQA~\cite{yang_hotpotqa_2018}, WebShop~\cite{yao_webshop_2023}, and Alfworld~\cite{shridhar_alfworld_2021} to the \ac{swe} domain. We introduce a repository-specific \ac{ctim} and investigate its performance on a subset of SWE-bench Verified~\cite{openai_swe_bench_verified_2024}. Our results show that this simple \ac{el} implementation does not generalize to the \ac{swe} setting and that the findings reported by~\citet{zhao_expel_2024} and~\citet{lingam_enhancing_2024} are thus limited to simpler environments with shorter trajectories. An extension of a general \ac{ctim} with repository-level \ac{ctim} or exemplar-based \ac{icl} does not suffice to amend this. Our investigations reveal noisy \ac{ctim} items as the likely culprit. We suggest removing the maximum size constraint from \ac{ctim} and instead focus on embedding-based retrieval of highly relevant \ac{ctim} items with respect to the problem statement. Furthermore, we suspect that the use of the~\ac{ctim} should not be limited to the initial context, and instead a subset of relevant items should be provided at each turn~\cite{su_learn-by-interact_2025}.

\section*{Limitations}
\label{sec:limitations}
\label{sec:limitations}
Regarding our dataset a key limitation is that we had to remove $\approx 10\%$ of samples from SWE-bench Verified due to defective environment setup scripts. We only removed these samples after validating these issues with the AutoCodeRover authors. Furthermore, while our ablation study and qualitative analysis hint at a potential path towards a concise and focused \ac{ctim} implementation that improves, rather than degrades performance via embedding-based retrieval of \ac{ctim} items, we do not investigate such an approach. Future work may also consider the re-use of self-reflections as \ac{ctim} items more explicitly. In our work we only implicitly distill knowledge from these in the success-failure tuple setting. 

\section*{Acknowledgements}
We acknowledge the support of this project by Jetbrains who kindly provided API credits for our experiments. We thank the AutoCodeRover~\cite{zhang_autocoderover_2024} authors for the helpful discussions and support in building on top of their research. Tobias Lindenbauer thanks Sophia Schwarz for her unwavering support throughout this project.
\bibliography{custom}

\appendix
\label{sec:appendix}

\section{Our Train and Test Sets Based on SWE-bench Verified}
\label{sec:appendix:dataset}
\begin{figure}[th] 
    \centering
    \includegraphics[width=0.5\textwidth]{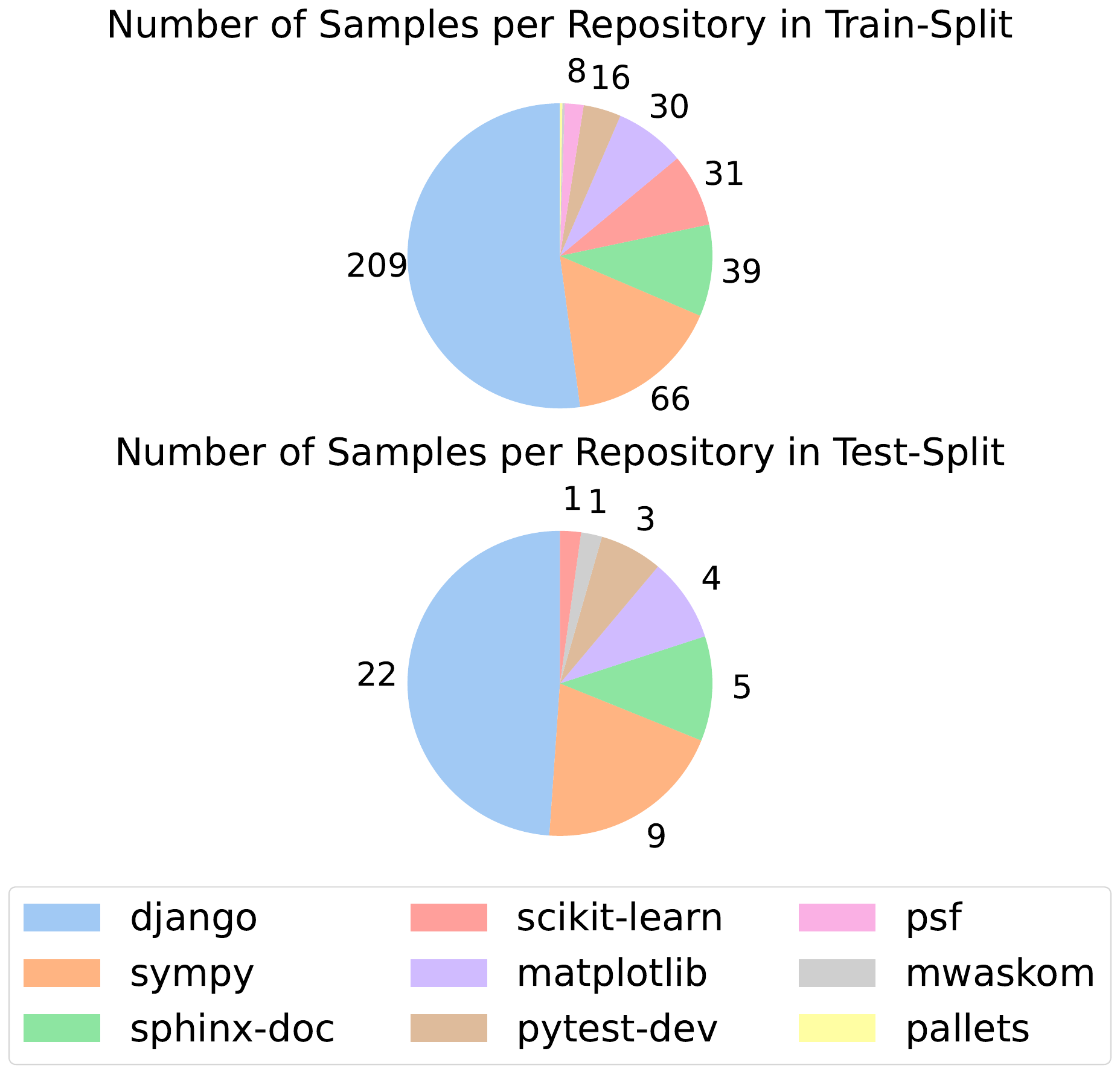} 
    \caption{The distribution of repositories across our train and test sets.}
    \label{fig:appendix:repository-distribution}
\end{figure}
\begin{figure}[th] 
    \centering
    \includegraphics[width=0.5\textwidth]{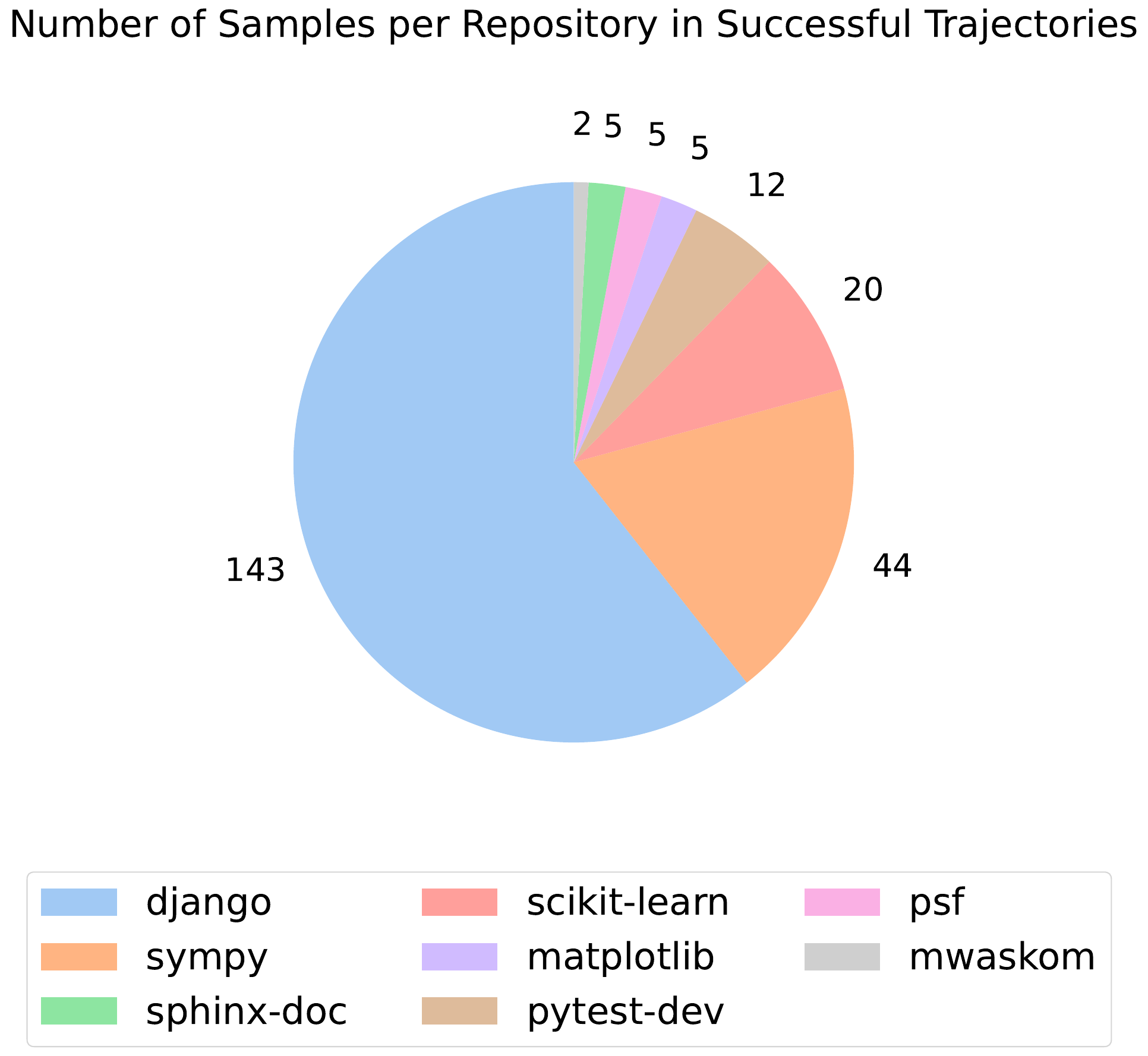} 
    \caption{The distribution of repositories across successful solved instanced by \acs{ctim}-Rover on our train split and by AutoCodeRover on SWE-bench Lite~\cite{jimenez_swe-bench_2023}. In total there are 236 solved instances based on which we create our \ac{ctim}.}
    \label{fig:appendix:successful-trajs:repository-distribution}
\end{figure}

We use the human-annotation data released with SWE-bench Verified~\cite{openai_swe_bench_verified_2024} for partitioning SWE-bench Verified into a train and test set. For this, we first investigate the statistical association of the \texttt{underspecified}, \texttt{false\_negative} and \texttt{difficulty} features from the annotation data with the outcome of an instance being solved by any competing system on SWE-bench Verified, to inform our decision across which fields to stratify during the dataset partitioning. We base this analysis on a snapshot of the SWE-bench Verified leader board from October 30, 2024. Table~\ref{tab:appendix:chi-square} presents the p-values for the investigated features. We find that the \texttt{difficulty} and \texttt{false\_negative} features are statistically significantly associated with the resolution of an instance at a significance level of $\alpha = 0.01$. These fields correspond to the subjective time required to resolve an instance by human annotators and whether the test suite mistakenly filters out successful solutions due to overly specific tests (e.g., requiring specific error messages), respectively. We did not find an \texttt{underspecifed} problem statement to statistically significantly affect instance resolution performance. We thus stratify across the statistically significant features to construct train and test sets that are representative of the original dataset across these success-related features.

In Figure~\ref{fig:appendix:repository-distribution}, we show the distribution of repositories across our train and test sets and observe that data distribution of repositories decently approximated across both the train and test set even without explicitly considering this in our stratification process. However, some outliers exist (e.g., ``psf'' or ``pallets''). For these all samples are contained in the train set. Our partitioning approach thus mostly maintains a repository overlap between train and test sets, a critical prerequisite for the cross-repository knowledge transfer of repository-level knowledge in our \ac{ctim} approach.

\begin{table*}[ht]
\centering
\begin{tabular}{lccc}
\hline
\textbf{Feature} & \textbf{p-value} & \textbf{$\chi^2$ statistic} & \textbf{Used For Stratification} \\
\hline
\texttt{underspecified} & 0.6293 & 0.2329 & \\
\texttt{false\_negative} & $6.3\cdot10^{-4}$ & 11.6946 & \checkmark \\
\texttt{difficulty} & $3.4\cdot10^{-4}$ & 18.5154 & \checkmark \\
\hline
\end{tabular}
\caption{Chi-square analysis of SWE-bench Verified annotations and instance resolution status.}
\label{tab:appendix:chi-square}
\end{table*}

\section{Extended Related Work}
\label{app:related-work}

\subsection{\acf{el}}
\label{app:related-work:se-agents}
One exciting advantage of \ac{el} and \ac{ctim} is that they allow agents to continually augment their knowledge of dynamic environments through \ac{icl}.
\ac{el} is loosely related to \ac{cl}. However, typical \ac{el} methods require updating a model's weights or architecture~\cite{shi_continual_2024}. This is often unfeasible and exposes the model to the risk of catastrophic forgetting~\cite{kirkpatrick_overcoming_2017}. While our implementation of \ac{el} did not improve baseline performance, the findings of~\citet{su_learn-by-interact_2025} support our finding that noisy context is the cause of the performance degradation and provide an actionable way forward for tackling this issue: embedding-based retrieval of \ac{ctim} items at every agent turn.

\section{Extended Results}
\label{sec:appendix:results}
To motivate the increased complexity of \ac{swe} compared to the domains investigated by ExpeL~\cite{zhao_expel_2024} we supplement our discussion of this matter in Section~\ref{sec:introduction} with statistics on turns taken and tokens consumed (Table~\ref{tab:appendix:trajectory-stats}. Note that the consumed tokens map to the context window at the final step in which the agent generates its patch.

\begin{table*}[t]
\centering
\small
\begin{tabular}{lrrrrrrrrr}
\hline
\multirow{2}{*}{\textbf{Configuration}} & \multicolumn{4}{c}{\textbf{Turns}} & \multicolumn{4}{c}{\textbf{Tokens}}  \\
\cline{2-9}
& \textbf{Median} & \textbf{Mean} & \textbf{Min} & \textbf{Max} & \textbf{Median} & \textbf{Mean} & \textbf{Min} & \textbf{Max} & \\
\hline
AutoCodeRover            & 9  & 11.16 & 4  & 20 & 10027 & 11414.02 & 3654 & 31408  \\
\acs{ctim}-Rover & 7  & 9.62  & 4  & 20 & 17807 & 17544.93 & 4962 & 36250  \\
~~~\acs{ctim} only             & 8  & 10 & 4  & 20 & 10724 & 13200.96 & 4438 & 46983  \\
~~~General \acs{ctim} only         & 9  & 10.31 & 4  & 20 & 9984  & 12272.22 & 3594 & 37785 \\
~~~Repo-level \acs{ctim} only            & 8  & 9.67  & 4  & 20 & 10130 & 12139.87 & 3953 & 38730 \\
~~~Exemplar only             & 8  & 9.2  & 4  & 20 & 15579 & 16143.47 & 3365 & 43799\\
\hline
\end{tabular}
\caption{Summary statistics for turn and token lengths on our test set ($n=45$) across different configurations. All numeric values are rounded to two decimal places.}
\label{tab:appendix:trajectory-stats}
\end{table*}

\section{\acf{ctim} Construction}
In this section we provide further details on the construction of our \acp{ctim}. In Section~\ref{fig:appendix:successful-trajs:repository-distribution} we discuss the distribution of the trajectories that serve as basis for \ac{kd} and \ac{ctim} construction in relation to our train and test set's data distributions. Then, we detail our prompt templates for \ac{ctim} creation in Section~\ref{ctim-prompts}. Finally, we provide sample \ac{ctim} items in Section~\ref{sample-ctim-items}.

\subsection{Distribution of Past Successful Trajectories For \acf{kd}}
\label{sec:appendix:training-trajectory-distr}
Compared to the distributions of our train and test sets in Figure~\ref{fig:appendix:repository-distribution} the available past successful trajectories for the actual \ac{ctim} construction are even more heavily skewed towards the repositories ``django'' and ``sympy''~(Figure~\ref{fig:appendix:successful-trajs:repository-distribution}). This is problematic for the general-level \ac{ctim} because it makes it challenging to distill knowledge that is generally relevant to \ac{swe}. To amend this, we oversample from all repositories except ``django'' in the ``sets of successful past trajectories'' \ac{kd} setting such that we show an evenly balanced number trajectories across ``django'' and all remaining repositories. For the repository-specific \ac{ctim} on the other hand, we expected the ``django'' repository to benefit from the additional knowledge in particular due to this heavy skew.

\subsection{Knowledge Distillation Prompts}
\label{ctim-prompts}
\definecolor{promptorangeaccent}{RGB}{252, 128, 29}
\definecolor{promptorange}{RGB}{254, 204, 165}

\definecolor{promptlavender}{RGB}{230, 230, 250}
\definecolor{promptlavenderaccent}{RGB}{99,99,224}

\definecolor{promptturqoise}{RGB}{165, 235, 225}
\definecolor{promptturqoiseaccent}{RGB}{26, 121, 105}

\definecolor{promptpinkaccent}{RGB}{255,49,140}
\definecolor{promptpink}{RGB}{255,205,228}

\begin{figure}[th]
    \centering
    \includegraphics[width=0.45\textwidth]{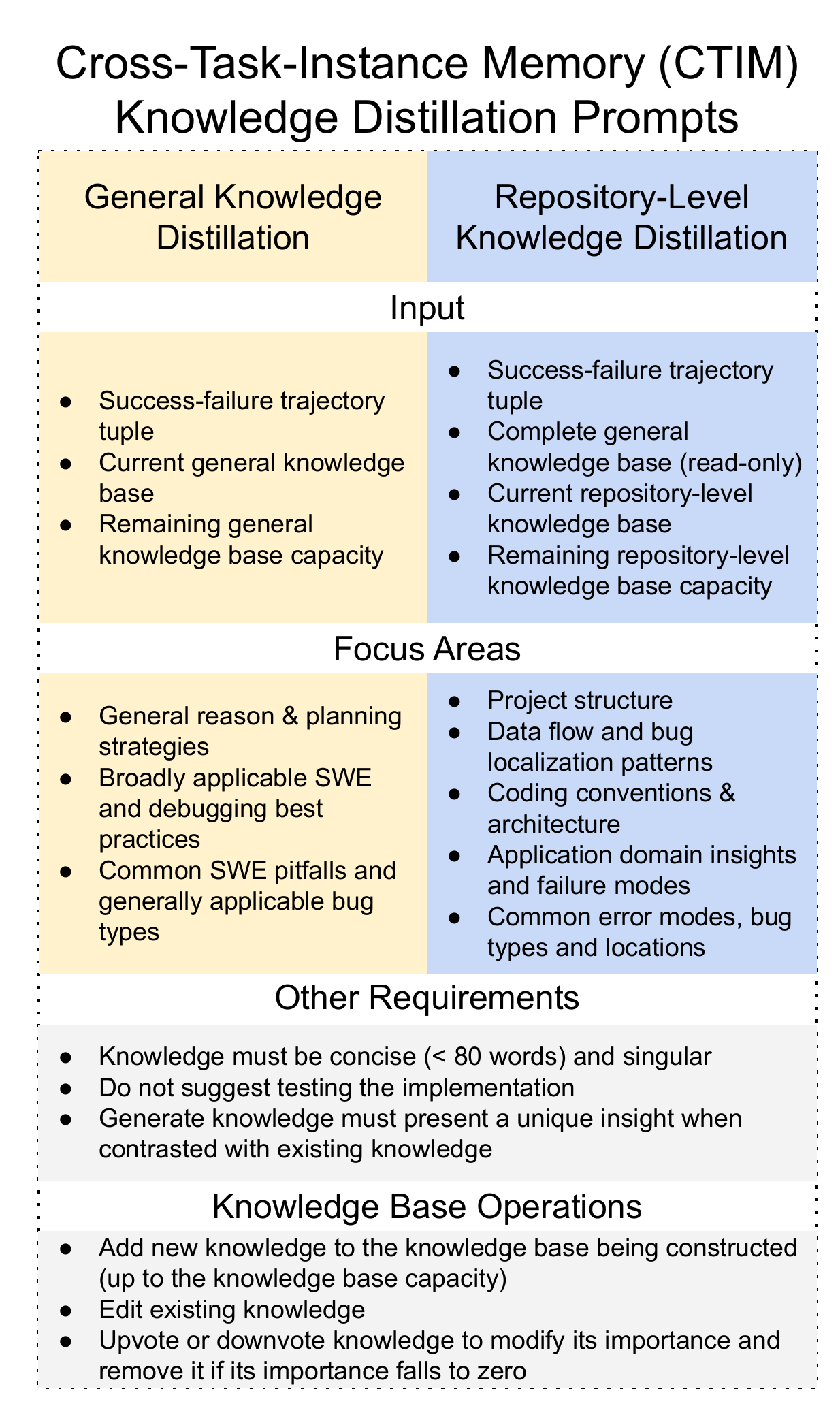}
    \caption{Key differences in our prompting strategies for general and repository-level \acf{kd}. The general \ac{kd} (left) captures broadly applicable software engineering principles, while the repository-level distillation (right) captures repository-specific patterns. In the repository-level \ac{kd}, we ensure that all repository-specific data originate from the same repository. In our implementation, we refer to knowledge items as "rules" in prompting templates, but conceptually they represent distilled knowledge.}
    \label{fig:appendix:kd-prompt}
\end{figure}

Figure~\ref{fig:appendix:kd-prompt} illustrates the prompt structure used in the success-failure tuple setting. Both the general and repository-level \ac{kd} approaches enforce a common structural framework for the distilled knowledge and support identical operations on their respective knowledge bases. However, they differ notably in terms of input configuration and focus: for general \ac{kd}, the model is prompted to extract broadly applicable software engineering best practices along with common error and bug types, whereas repository-level \ac{kd} explicitly targets repository-specific information—such as the repository’s structure, data flow patterns, and localized characteristics. In the repository-level phase, the prompt also incorporates the previously extracted high-level general knowledge as a reference, ensuring that the new insights provide a distinct, fresh perspective. Moreover, the distilling LLM is restricted to modifying only the repository-specific knowledge base for the repository currently being processed, even though the general knowledge base remains visible solely for comparison purposes. For further details please refer to Figures~\ref{fig:prompt:system} to~\ref{fig:prompt:repo-level-pair-p2} detailing our prompts below.

\label{sec:appendix:prompt-template}

\begin{figure*}[t]
\begin{tcolorbox}[
  colback=promptlavender,
  colframe=promptlavenderaccent,
  title=Knowledge Distillation System Prompt, width=\textwidth
]
\begin{small}
\begin{verbatim}
You are an advanced reasoning agent that can ADD, EDIT, UPVOTE or DOWNVOTE rules from an existing rule 
set, which is constructed by reflecting on and critiquing past successful task trajectories.
\end{verbatim}
\end{small}
\end{tcolorbox}
\caption{The system prompt we use for both \ac{kd} phases and all \ac{kd} settings.}
\label{fig:prompt:system}
\end{figure*}

\begin{figure*}[t]
\begin{tcolorbox}[
  colback=promptlavender,
  colframe=promptlavenderaccent,
  title=\ac{ctim} Capacity Warning Prompt, width=\textwidth
]
\begin{small}
\begin{verbatim}
You have reached the maximum ruleset size of {ruleset_cap}. The ADD  operation is now INVALID. To 
reduce the ruleset size, prune low-utility rules that overlap with others by performing the 
DOWNVOTE operation on them.
\end{verbatim}
\end{small}
\end{tcolorbox}
\caption{The \ac{ctim} capacity warning prompt we use in both \ac{kd} phases and all \ac{kd} settings.}
\label{fig:prompt:ctim-cap}
\end{figure*}

\begin{figure*}[t]
\begin{tcolorbox}[
  colback=promptlavender,
  colframe=promptlavenderaccent,
  title=\ac{ctim} Capacity Information Prompt, width=\textwidth
]
\begin{small}
\begin{verbatim}
You may add up to {remaining_slots} more rules to the ruleset before reaching the maximum 
of {ruleset_cap} rules.
\end{verbatim}
\end{small}
\end{tcolorbox}
\caption{The \ac{ctim} capacity information prompt we use in both \ac{kd} phases and all \ac{kd} settings.}
\label{fig:prompt:ctim-info}
\end{figure*}

\begin{figure*}[t]
\begin{tcolorbox}[
  colback=promptorange,
  colframe=promptorangeaccent,
  title=\ac{ctim} Operations Prompt - General \ac{kd} (Phase 1), width=\textwidth
]
\begin{small}
\begin{verbatim}
Provide the operations as a list
containing JSON objects of the following schema:
{{
"operation_type": {{"enum": ["ADD", "EDIT", "UPVOTE", "DOWNVOTE"]}},
"rule_id": {{"type": "integer"}},
"rule_content": {{"type": "string"}}
}}
The "operation_type" field specifies the type of operation to perform on the rule with the given 
"rule_id". The "rule_id" must be an integer identifying a rule in the current 
ruleset{ruleset_indices_hint}. If you are adding or editing a rule, additionally provide 
the "rule_content" field with the new content of the rule.

Here is an example of a valid response:
{{"operations":
    [{{
        "operation_type": "ADD",
        "rule_content": <Extracted insight, knowledge, tip or rule>
    }},
    {{
        "operation_type": "DOWNVOTE",
        "rule_id": <Integer identifying an EXISTING rule that is contradictory to
                    another rule, this sample or too similar to another rule>
    }},
    {{
        "operation_type": "EDIT",
        "rule_id": <Integer identifying an EXISTING rule>,
        "rule_content": <Extracted insight, knowledge, tip or rule to update and
                         enhance the EXISTING rule with>
    }}]
}}
\end{verbatim}
\end{small}
\end{tcolorbox}
\caption{The \ac{ctim} operations prompt we use in the high level \ac{kd} phase and both its \ac{kd} settings.}
\label{fig:prompt:ctim-operations-p1}
\end{figure*}

\begin{figure*}[t]
\begin{tcolorbox}[
  colback=promptorange,
  colframe=promptorangeaccent,
  title=\ac{ctim} Operations Prompt - General \ac{kd} (Phase 1) continued, width=\textwidth
]
\begin{small}
\begin{verbatim}
Do not mention the trajectories or their ids explicitly in your responses. Do not reference specific 
file, class, function or variable names to ensure that your ruleset is general and transferable to
other task instances and repositories. You can use any of the valid operation types multiple times. 

Each existing rule can be modified only once. 
The following operations are valid:
- UPVOTE an EXISTING rule if it is strongly relevant in your current context and
  trajectories. Valid fields: [operation_type, rule_id]
- DOWNVOTE an EXISTING rule if the rule contradicts your current context and
  trajectories or is similar to or a duplicate
of another existing rule. Make use of this operation to achieve a concise ruleset
that is relevant across repositories and task instances.
If you downvote a rule often enough it will be removed from the ruleset. Valid
fields: [operation_type, rule_id]
- EDIT an EXISTING rule if it is not general enough or could be enhanced given your
  current context by rewriting, adding or removing content. Valid fields:
  [operation_type, rule_id, rule_content]
- ADD a NEW rule if you identified insights that are generally applicable and
  transferable to other task instances. Make sure that the new rule is distinct
from existing rules. Valid fields: [operation_type, rule_content]

Key requirements:
- The only operation that is valid on rules that do not yet exist is ADD.
- If you have reached the maximum ruleset size, you must not add any new rules.
  Instead, you must edit existing rules or upvote/downvote existing rules.
- You may provide between 1 and 4 operations.
\end{verbatim}
\end{small}
\end{tcolorbox}
\caption{The \ac{ctim} operations prompt we use in the high level \ac{kd} phase and both its \ac{kd} settings continued.}
\label{fig:prompt:ctim-operations-p1-2}
\end{figure*}

\begin{figure*}[t]
\begin{tcolorbox}[
  colback=promptorange,
  colframe=promptorangeaccent,
  title=Success-Failure Trajectory \ac{kd} Setting - General \ac{kd} (Phase 1), width=\textwidth
]
\begin{small}
\begin{verbatim}
You are given a set of successful task trajectories that relate to fixing bugs in open-source code 
repositories. During these trajectories you correctly identified the location of the buggy code, 
wrote a patch which fixed the bug in the code and passed all test cases, meaning you also didn't in 
introduce any new bugs.

Below follow the past successful task trajectories. The set of trajectories is delimited by the 
<PAST_SUCCESSFUL_TRAJECTORIES> and </PAST_SUCCESSFUL_TRAJECTORIES> tags. Each trajectory is 
wrapped by the <TRAJECTORY-i> and </TRAJECTORY-i> tags, where i identifies the i-th trajectory 
in the set below:
<PAST_SUCCESSFUL_TRAJECTORIES>
{past_successful_trajectories}
</PAST_SUCCESSFUL_TRAJECTORIES>

Next, follow a set of rules that you have extracted so far. The ruleset is limited to {ruleset_cap} 
rules. Any rules beyond {ruleset_cap} rules will be ignored:
{current_ruleset}
{remaining_slots_information}

By examining the successful trajectories, and the existing rules above you should update the 
existing ruleset by adding, editing, upvoting or downvoting rules. The resulting ruleset must 
consist of high-level knowledge, insights or tips that are generally applicable, covering 
the following aspects:
1. Reasoning and planning strategies that serve as guiding signals for future task attempts, 
especially with respect to identifying the locations of buggy code effectively.
2. Coding practices, patterns, and idioms that are generally applicable to writing high-quality, 
staff senior level code, to fix bugs.
3. Common pitfalls and error patterns in software engineering that are relevant to identifying 
and fixing buggy code.

\end{verbatim}
\end{small}
\end{tcolorbox}
\caption{The prompt describing the success-failure trajectory pair \ac{kd} setting for the high level \ac{kd} phase.}
\label{fig:prompt:sf-p1}
\end{figure*}

\begin{figure*}[t]
\begin{tcolorbox}[
  colback=promptorange,
  colframe=promptorangeaccent,
  title=Success-Failure Trajectory \ac{kd} Setting - General \ac{kd} (Phase 1) continued, width=\textwidth
]
\begin{small}
\begin{verbatim}
Key requirements for rules:
- DO NOT suggest testing the implementation. The agent using your ruleset is UNABLE to test 
its implementation. It must generate a correct patch on the first attempt.
- Generated rules must be concise (less than 80 words) and should be focused on a single, 
specific aspect or insight.
- Generated rules must be unique with respect to other, existing rules and contribute a new, 
unique piece of information, knowledge or perspective.

This ruleset should serve as the basis for guiding future task attempts in locating and 
fixing bugs to a successful completion and empower the agent to improve its planning, 
reasoning, coding skills, bug localization skills.
\end{verbatim}
\end{small}
\end{tcolorbox}
\caption{The prompt describing the success-failure trajectory pair \ac{kd} setting for the high level \ac{kd} phase continued.}
\label{fig:prompt:sf-p1-2}
\end{figure*}

\begin{figure*}[t]
\begin{tcolorbox}[
  colback=promptorange,
  colframe=promptorangeaccent,
  title=Sets of Successful Trajectories \ac{kd} Setting - General \ac{kd} (Phase 1), width=\textwidth
]
\begin{small}
\begin{verbatim}
Below you will find multiple past attempts at fixing a bug in an open-source code repository. 
The first few trajectories show failed attempts, the last trajectory shows a successful bug fix.
All attempts are related to fixing the same bug in the same codebase. Compare and contrast the 
successful and failed attempts to understand why the initial attempts failed and which change 
in the reasoning, planning, coding or bug localization strategy could have led to a correct 
patch generation in the first attempt. Consider the self-reflections that took place between 
the failed attempts to understand which changes were made in the reasoning, planning, coding 
or bug localization strategy that led to the bug being fixed in the last trajectory.

Below follow the task attempts denoted by <FAILED_TASK_ATTEMPT-i> and </FAILED_TASK_ATTEMPT-i> tags 
where i identifies the i-th failed attempt and the successful task attempt is denoted by the 
<SUCCESSFUL_TASK_ATTEMPT> and </SUCCESSFUL_TASK_ATTEMPT> tags. Only failed task attempts contain 
a self-reflection:
{success_failure_trajectory}

Next, follow a set of rules that you have extracted so far. The ruleset is limited to {ruleset_cap} 
rules. Any rules beyond {ruleset_cap} rules will be ignored:
{current_ruleset}

{remaining_slots_information}

By examining and comparing the successful and failed attempts, and the existing rules above you 
should update the existing ruleset by adding, editing, upvoting or downvoting rules. The resulting 
ruleset must consist of high-level knowledge, insights or tips that are generally applicable, 
covering the following aspects:
1. Reasoning and planning strategies that serve as guiding signals for future task attempts, 
especially with respect to entifying the locations of buggy code effectively.
2. Coding practices, patterns, and idioms that are generally applicable to writing high-quality, 
staff senior level code, to fix bugs.
3. Common pitfalls and error patterns in software engineering that are relevant to identifying 
and fixing buggy code.

Key requirements for rules:
- DO NOT suggest testing the implementation. The agent using your ruleset is UNABLE to test 
its implementation. It must generate a correct patch on the first attempt.
- DO NOT suggest reflecting on a past trajectory or attempt. The agent using your ruleset 
is UNABLE to reflect on a past trajectory or attempt. It must generate a correct patch on 
the first attempt.
\end{verbatim}
\end{small}
\end{tcolorbox}
\caption{The prompt describing the sets of successful trajectories \ac{kd} setting for the high level \ac{kd} phase.}
\label{fig:prompt:sets-p1}
\end{figure*}

\begin{figure*}[t]
\begin{tcolorbox}[
  colback=promptorange,
  colframe=promptorangeaccent,
  title=Sets of Successful Trajectories \ac{kd} Setting - General \ac{kd} (Phase 1) continued, width=\textwidth
]
\begin{small}
\begin{verbatim}
- Generated rules must be concise (less than 80 words) and should be focused on a single, 
specific aspect or insight.
- Generated rules must be unique with respect to other, existing rules and contribute a 
new, unique piece of information, knowledge or perspective.

This ruleset should serve as the basis for guiding future task attempts in locating and fixing 
bugs to a successful completion.
It should empower the agent to improve its planning, reasoning, coding, and bug localization 
skills.

\end{verbatim}
\end{small}
\end{tcolorbox}
\caption{The prompt describing the sets of successful trajectories \ac{kd} setting for the high level \ac{kd} phase continued.}
\label{fig:prompt:sets-p1-2}
\end{figure*}

\begin{figure*}[t]
\begin{tcolorbox}[
  colback=promptpink,
  colframe=promptpinkaccent,
  title=\ac{ctim} Operations Prompt - Repository-Level \ac{kd} (Phase 2), width=\textwidth
]
\begin{small}
\begin{verbatim}
Provide the operations as a list containing JSON objects of the following schema:
{{
"operation_type": {{"enum": ["ADD", "EDIT", "UPVOTE", "DOWNVOTE"]}},
"rule_id": {{"type": "integer"}},
"rule_content": {{"type": "string"}},
"knowledge_type": {{"enum": ["repository_structure", "architectural_pattern",
"coding_convention", "error_pattern",
"application_domain"]}}
}}
The "operation_type" field specifies the type of operation to perform on the rule with the given 
"rule_id". The "rule_id" must be an integer identifying a rule in the current 
ruleset{ruleset_indices_hint}. If you are adding or editing a rule, additionally 
provide the "rule_content" field with the new content of the rule. If you are adding 
a rule, you must also specify the "knowledge_type" of the rule.

Here is an example of a valid response:
{{"operations":
    [{{
        "operation_type": "ADD",
        "rule_content": <Extracted insight, knowledge, tip or rule>,
        "knowledge_type": "error_pattern"
    }},
    {{
        "operation_type": "ADD",
        "rule_content": <Knowledge about the application domain of the project and typical
                         edge cases resulting from this.>
        "knowledge_type": "application_domain"
    }},
    {{
        "operation_type": "DOWNVOTE",
        "rule_id": <Integer identifying an EXISTING rule that is contradictory to another
                    rule, this sample or too similar to another rule>
    }},
    {{
        "operation_type": "EDIT",
        "rule_id": <Integer identifying an EXISTING rule>,
        "rule_content": <Extracted insight, knowledge, tip or rule to update and enhance
                         the EXISTING rule with>
    }}]
}}
\end{verbatim}
\end{small}
\end{tcolorbox}
\caption{The \ac{ctim} operations prompt we use in the repository-level \ac{kd} phase and both its \ac{kd} settings.}
\label{fig:prompt:ctim-repo-operations-p1}
\end{figure*}

\begin{figure*}[t]
\begin{tcolorbox}[
  colback=promptpink,
  colframe=promptpinkaccent,
  title=\ac{ctim} Operations Prompt - Repository-Level \ac{kd} (Phase 2) continued, width=\textwidth
]
\begin{small}
\begin{verbatim}
Do not mention the trajectories or their ids explicitly in your responses. You may reference specific 
file, class, function names, but keep in mind that the repository evolves over time and files, 
classes or functions may be renamed, removed or refactored. You can use any of the valid operation 
types multiple times. Each existing rule can be modified only once. The following operations 
are valid:

- UPVOTE an EXISTING rule if it is strongly relevant in your current context and trajectories.
  Valid fields: [operation_type, rule_id]
- DOWNVOTE an EXISTING rule if the rule contradicts your current context and trajectories or it
  is similar to or a duplicate of another existing rule (including general purpose rules).
  Make use of this operation to achieve a concise ruleset that is relevant across 
  repositories and task instances. If you downvote a rule often enough it will be 
  removed from the ruleset. Valid fields: [operation_type, rule_id]
- EDIT an EXISTING rule if it is not general enough or could be enhanced given your current
  context by rewriting, adding or removing content. Valid fields: [operation_type, rule_id,
  rule_content]
- ADD a NEW rule if you identified insights that are generally applicable and potentially
  beneficial to other task instances in the same repository. Make sure that the new rule is
  unique. Valid fields: [operation_type, rule_content, knowledge_type]

Key requirements:
- The only operation that is valid on rules that do not yet exist is ADD.
- If you have reached the maximum ruleset size, you must not add any new rules. Instead, you
  must edit existing rules or upvote/downvote existing rules.
- You may provide between 1 and 4 operations.
\end{verbatim}
\end{small}
\end{tcolorbox}
\caption{The \ac{ctim} operations prompt we use in the repository-level \ac{kd} phase and both its \ac{kd} settings. Continued.}
\label{fig:prompt:ctim-repo-operations-p2}
\end{figure*}

\begin{figure*}[t]
\begin{tcolorbox}[
  colback=promptpink,
  colframe=promptpinkaccent,
  title=Success-Failure Trajectory \ac{kd} Setting - Repository-Level \ac{kd} (Phase 2), width=\textwidth
]
\begin{small}
\begin{verbatim}
You are given a set of successful task trajectories that relate to fixing issues the real-world 
repository '{repository_name}'. During these trajectories you correctly identified the location 
of the buggy code, wrote a patch which fixed the bug in the code and passed all test cases, 
meaning you also didn't in introduce any new bugs. Due to the natural evolution of software over 
time the state of the repository when you carried out the tasks in the example trajectories below 
may differ slightly. You might encounter differences with respect to the project structure, and 
file, class, method or variable names. If you encounter conflicting information, do not record 
any rules regarding the conflicting elements.

Below follow the past successful task trajectories. The set of trajectories is delimited by the 
<PAST_SUCCESSFUL_TRAJECTORIES> and </PAST_SUCCESSFUL_TRAJECTORIES> tags. Each trajectory is 
wrapped by the <TRAJECTORY-i> and </TRAJECTORY-i> tags, where i identifies the i-th trajectory 
in the set below:
<PAST_SUCCESSFUL_TRAJECTORIES>
{past_successful_trajectories}
</PAST_SUCCESSFUL_TRAJECTORIES>

Next, follows the frozen set of high-level, general purpose rules that you have extracted previously. 
These rules are READ-only, you must not perform any operations on them. You may refer to these rules 
directly in the repository level rules as 'GENERAL PURPOSE RULE-i' to highlight their specific 
application, knowledge gaps or discrepancies with respect to the current repository:
{general_ruleset}
\end{verbatim}
\end{small}
\end{tcolorbox}
\caption{The prompt describing the sets of successful trajectories \ac{kd} setting for the repository-level \ac{kd} phase.}
\label{fig:prompt:repo-level-sets-p1}
\end{figure*}

\begin{figure*}[t]
\begin{tcolorbox}[
  colback=promptpink,
  colframe=promptpinkaccent,
  title=Success-Failure Trajectory \ac{kd} Setting - Repository-Level \ac{kd} (Phase 2) continued, width=\textwidth
]
\begin{small}
\begin{verbatim}


Below follows the modifiable set of repository-level rules that you have extracted so far. The 
repository-level ruleset is limited to {ruleset_cap} rules. Any rules beyond {ruleset_cap} rules 
will be ignored:
{current_repository_level_ruleset}

By examining the successful trajectories, and the existing general purpose and repository-level 
rules above you should update the repository-level ruleset by adding, editing, upvoting or downvoting 
repository-level rules. The resulting ruleset must consist of repository-specific knowledge, 
insights or tips that are unique to this codebase and provide new insights that are distinct from the 
general purpose rules. Repository-level rules may cover the following aspects:
1. Repository-level bug localization and environment exploration patterns that help locate relevant 
code sections quickly, including key file locations, module relationships.
2. Repository-level coding conventions, architectural principles, design patterns, and 
implementation approaches that are consistently used across the codebase and should be followed 
when making changes.
3. Repository-level error or exception handling strategies, including custom errors or exceptions
4. The application domain of the project (e.g., Does the software handle images or text and what 
kind? Is it a command line application or does it have a GUI? Does it handle HTTP requests? Is it a 
highly technical, mathematical application?)
5. Common edge cases or failure modes related to the project's specific application domain. What 
are common errors or potential pitfalls in these application domains?).

Key requirements for rules:
- DO NOT suggest testing the implementation. The agent must generate correct patches on the first 
attempt by leveraging general and repository-specific rules identified above.
- Generated rules must be concise (less than 80 words) and should be focused on a single, specific 
aspect or insight.
- Generated rules must be unique with respect to other, existing rules and contribute a new, unique 
piece of information, knowledge or perspective.

This ruleset serves as the basis for guiding future task attempts within this repository in locating 
and fixing bugs to a successful completion. It should empower the agent to improve its planning, 
reasoning, coding, and bug localization skills.
    
{remaining_slots_information}
\end{verbatim}
\end{small}
\end{tcolorbox}
\caption{The prompt describing the sets of successful trajectories \ac{kd} setting for the repository-level \ac{kd} phase. Continued.}
\label{fig:prompt:repo-level-sets-p2}
\end{figure*}

\begin{figure*}[t]
\begin{tcolorbox}[
  colback=promptpink,
  colframe=promptpinkaccent,
  title=Success-Failure Trajectory \ac{kd} Setting - Repository-Level \ac{kd} (Phase 2), width=\textwidth
]
\begin{small}
\begin{verbatim}
Below you will find multiple past attempts at fixing a bug in an open-source code repository. The 
first few trajectories show failed attempts, the last trajectory shows a successful bug fix.
All attempts are related to fixing the same bug in the same codebase. Compare and contrast the 
successful and failed attempts to understand why the initial attempts failed and which change in 
the reasoning, planning, coding or bug localization strategy could have led to a correct patch 
generation in the first attempt.Consider the self-reflections that took place between the failed 
attempts to understand which changes were made in the reasoning, planning, coding or bug localization 
strategy that led to the bug being fixed in the last trajectory.

Below follow the task attempts denoted by <FAILED_TASK_ATTEMPT-i> and </FAILED_TASK_ATTEMPT-i> tags 
where i identifies the i-th failed attempt and the successful task attempt is denoted by the 
<SUCCESSFUL_TASK_ATTEMPT> and </SUCCESSFUL_TASK_ATTEMPT> tags. Only failed task attempts contain a 
self-reflection:
{success_failure_trajectory}

Next, follows the frozen set of high-level, general purpose rules that you have extracted previously. 
These rules are READ-only, you must not perform any operations on them. You may refer to these rules 


\end{verbatim}
\end{small}
\end{tcolorbox}
\caption{The prompt describing the success-failure trajectory pair \ac{kd} setting for the repository-level \ac{kd} phase.}
\label{fig:prompt:repo-level-pair-p1}
\end{figure*}

\begin{figure*}[t]
\begin{tcolorbox}[
  colback=promptpink,
  colframe=promptpinkaccent,
  title=Success-Failure Trajectory \ac{kd} Setting - Repository-Level \ac{kd} (Phase 2) continued, width=\textwidth
]
\begin{small}
\begin{verbatim}
directly in the repository level rules as 'GENERAL PURPOSE RULE-i' to highlight their specific 
application, knowledge gaps or discrepancies with respect to the current repository:
{general_ruleset}

Below follows the modifiable set of repository-level rules that you have extracted so far. The 
repository-level ruleset is limited to {ruleset_cap} rules. Any rules beyond {ruleset_cap} rules
will be ignored:
{current_repository_level_ruleset}
\end{verbatim}
\end{small}
\end{tcolorbox}
\caption{The prompt describing the success-failure trajectory pair \ac{kd} setting for the repository-level \ac{kd} phase. Continued.}
\label{fig:prompt:repo-level-pair-p2}
\end{figure*}

\subsection{Sample \ac{ctim} Items}
\label{sample-ctim-items}
In this section we provide sample general and repository-level \ac{ctim} items. We provide four randomly selected general (Figure~\ref{fig:ctim-general}) and django repository-level (Figure~\ref{fig:ctim-django}) items. Additionally, we provide all repository-level \ac{ctim} items for psf (Figure~\ref{fig:ctim-psf}), an underrepresented repository.

\begin{figure*}[t]
\begin{tcolorbox}[
  colback=promptturqoise,
  colframe=promptturqoiseaccent,
  title=Sample General \ac{ctim} Items, width=\textwidth
]
\begin{small}
\begin{verbatim}
- Perform targeted input validations, ensuring each parameter or feature aligns with immediate needs 
and preventing unwanted callability, type-mismatch, or boundary issues.

- Examine error messages to locate the failing logic. Also confirm if the framework\u2019s checks 
might be incomplete or incorrectly flag valid usage, especially for advanced lookups or 
edge cases.

- Always confirm that referenced methods or variables exist, are spelled correctly, remain valid, 
and that decorators or partials do not obscure them.

- Focus changes on the minimal relevant locations, referencing existing methods or design 
patterns to maintain consistency, reduce duplication, and lower risk of new bugs.
\end{verbatim}
\end{small}
\end{tcolorbox}
\caption{Four random general \ac{ctim} samples.}
\label{fig:ctim-general}
\end{figure*}

\begin{figure*}[t]
\begin{tcolorbox}[
  colback=promptturqoise,
  colframe=promptturqoiseaccent,
  title=Sample Django Repository-Level \ac{ctim} Items, width=\textwidth
]
\begin{small}
\begin{verbatim}
- When refactoring special-case or zero-quantity paths in the app\u2019s code 
(like max_post_process_passes=0), skip irrelevant steps entirely to avoid referencing uninitialized 
variables. If the field or setting indicates no passes or empty states, ensure the logic short-circuits 
properly. This avoids spurious errors from referencing variables that never get assigned.

- When unregistering or registering custom lookups in RegisterLookupMixin, always call 
_clear_cached_lookups afterward to avoid stale lookup references and maintain consistency 
with register_lookup.

- When retrieving fields from database insert operations using returning_fields, ensure 
from_db_value or equivalent logic is consistently applied, matching standard retrieval. 
This prevents raw values from bypassing normal conversions, especially for custom fields 
that rely on from_db_value to transform them into appropriate Python objects.

- When customizing admin logic in Django, including catch-all or fallback views, confirm usage of 
request.path vs request.path_info. request.path preserves the script name prefix required in certain 
redirects, while request.path_info omits it. Ensure to separate resolution from the final redirect
to keep path_info clean while preserving the prefix in the final URL, preventing forced script
name issues. Additionally, ensure admin checks referencing placeholders or fields include 
the actual field name in error messages for clarity.
\end{verbatim}
\end{small}
\end{tcolorbox}
\caption{Four random django \ac{ctim} samples.}
\label{fig:ctim-django}
\end{figure*}

\begin{figure*}[t]
\begin{tcolorbox}[
  colback=promptturqoise,
  colframe=promptturqoiseaccent,
  title=All psf Repository-Level \ac{ctim} Items, width=\textwidth
]
\begin{small}
\begin{verbatim}
- In 'prepare_headers' (requests/models.py::PreparedRequest), headers with a None value become 
the literal 'None' string if not filtered out. Always remove such keys to avoid invalid headers.


- When unregistering or registering custom lookups in RegisterLookupMixin, always 
call _clear_cached_lookups afterward to avoid stale lookup references and maintain consistency 
with register_lookup.

- When retrieving fields from database insert operations using returning_fields, ensure 
from_db_value or equivalent logic is consistently applied, matching standard retrieval. 
This prevents raw values from bypassing normal conversions, especially for custom fields 
that rely on from_db_value to transform them into appropriate Python objects.

- When customizing admin logic in Django, including catch-all or fallback views, confirm usage of 
request.path vs request.path_info. request.path preserves the script name prefix required in certain 
redirects, while request.path_info omits it. Ensure to separate resolution from the final redirect
to keep path_info clean while preserving the prefix in the final URL, preventing forced script
name issues. Additionally, ensure admin checks referencing placeholders or fields include 
the actual field name in error messages for clarity.
\end{verbatim}
\end{small}
\end{tcolorbox}
\caption{All psf repository-level \ac{ctim} samples.}
\label{fig:ctim-psf}
\end{figure*}

\end{document}